# A Framework for Developing Real-Time OLAP algorithm using Multi-core processing and GPU: *Heterogeneous Computing*

**H I Alzeini[1,4], Sh A Hameed[2] and M H Habaebi[3]**
Electrical and Computer Department, IIUM, PO: 53100, Jalan Gombak, Kuala-Lumpur, Malaysia

E-mail: [1] h.alzeini@hotmail.com, [2] shihab@iium.edu.my and [3] habaebi@iium.edu.my

**Abstract.** The overwhelmingly increasing amount of stored data has spurred researchers seeking different methods in order to optimally take advantage of it which mostly have faced a response time problem as a result of this enormous size of data. Most of solutions have suggested materialization as a favourite solution. However, such a solution cannot attain Real-Time answers anyhow. In this paper we propose a framework illustrating the barriers and suggested solutions in the way of achieving Real-Time OLAP answers that are significantly used in decision support systems and data warehouses.

## 1. Introduction

This Online analytical processing (OLAP) [1] is a category of software technology that enables analysts, managers and executives to gain insight into data through fast, consistent, interactive access to a wide variety of possible views of information that has been transformed from raw data to reflect the real dimensionality of the enterprise as understood by the user. OLAP functionality is characterized by dynamic multidimensional analysis of consolidated enterprise data supporting end user analytical and navigational activities including calculations and modelling applied across dimensions, through hierarchies and/or across members, trend analysis over sequential time periods, slicing subsets for on-screen viewing, drill-down to deeper levels of consolidation, rotation to new dimensional comparisons in the viewing area …etc. OLAP works with data warehouses that have sizes of not only terabytes but also petabytes which results in slower response time to answer queries. Consequently, processors have to tackle billions of rows. The issue has been intensively addressed and we would say that the most of solutions have relentlessly gone to materialization direction. Materialization by definition means pre-fetching data and pre-computing prospective queries – which are usually expected – load these answers with their features into the main memory. Subsequently, any coming query will find the answer waiting for it. This technique has been widely used and improved as it implies other problem such as the main memory occupation and excluding latest data. The latter issue has caused a significant debate, that is, by materializing result the system will present static answers that often do not take into account the latest updates of the warehouse. Warehouse usually is being updated periodically (e.g. daily, weekly, monthly … etc.). Therefore, pre-fetching and pre-computing are performed periodically either which would necessarily lead to some answers that do not include current time interval updates (non Real-Time answers). Thus, the Real-Time OLAP answers issues is about if the latest updates are included in the answers or not, rather than the absolute answer's time values – which varies dramatically according to several factors (e.g. query size, answer size, processing capabilities, distribution issues … etc.). Other minor solutions have been suggested will be elaborated in the following sections. However, all the materialization solutions cannot meet the Real-Time requirements which motivated other researchers to focus on different direction recently; how to optimally exploit processing facilities in order to function them to immediately answer queries. Such an approach, of course, needs powerful processing capabilities to process millions, and sometimes billions, of rows in few seconds to answer each query instantaneously. Recent suggestions included multicore processing, distributed processing as well as GPU utilizing.

---
[4] To whom any correspondence should be addressed.

The following sections of this paper are organized as follow: related work has been presented and briefly criticized. Then, the possible solution has been presented. Afterwards, our chosen suggestion has been shown in addition to the expected outcomes. Finally, a summary and future research directions.

**2. Materialization**

We can categorize the literature (to optimize materialization objective) into two main areas: how to optimize materialized views (sub-cube) construction and how to optimize materialized views distribution over multi-nodes systems.

*2.1. Sub-Cube Construction*

A dimension-oriented storage was one of the suggested solutions that have been proposed to minimize the cube aggregation time [2]. Z. Jing-hua et al have presented that the traditional row-oriented storage restricting the performance, accordingly; a new parallel aggregation mechanism utilizing MapReduce jobs [3] and HBase mechanism have been implemented and a performance improvement has been demonstrated. However, using HBase is restricted to several factors, such as the size of data must be enormous enough to achieve the efficiency (e.g. hundreds of millions or billions row) otherwise (e.g. few thousand/million rows) the traditional RDBMS is better choice due to the fact that HBase is efficient if we have 5 or more nodes at least. Furthermore, in order to use HBase we should tradeoff between extra features of RDBMS (e.g. transaction, typed columns, secondary indexes … etc.) and Hbase features. Third and most importantly; RDBMS cannot simply be ported to HBase by just replacing a JDBC driver, the operation require a full redesign cycle [5]. In spite of these limitations, one can benefit from [2] approach in parallelism part, using MapReduce [3] can be very helpful in developing a distributed computing system.

*2.2. Distributed OLAP*

Many researches have investigated views materialization over distributed processing systems as one of the potential solution aiming at reducing processing time consuming by OLAP queries. One of the resent researches for S. Sen et al [6] has focused on presenting new architecture to maintain materialized views for cloud computing environment. Considering the subtle differences between them; the proposed architecture is more pertained to distributed systems rather than the cloud computing.

A different infrastructure has been introduced by W. Li et al [7] with different utilization (distributed databases). In this paper there is single frontend database server and multiple backend databases. The materialized views (MQTs) have been used and the main contribution is the Data Placement Advisor (DPA - is a replacement of MQTA that had been designed to deal with standalone database server in which MQTs are placed on the same server with the base tables) that recommend MQTs and allocation mechanisms in order to minimize the response time of the queries. The new DPA build recommendations according to the analytic results of the MQTA and recommend whether to place MQTs at the backend or at the frontend based on the response time. A significant approach has been proposed in [8] defining the policies of OLAP queries caching in internet proxies, the solution suggests that we can utilize the LAN proxy by locating the materialized views in the proxy server in order to achieve the same purpose; reducing the response time for OLAP queries from scattered private LANs. The research suggests a mechanism to estimate access costs in order to decide whether the query will be satisfied locally (at the proxy) or at the central DW. An analytical cost model has been derived in order to quantify the benefits, a cache algorithm (VHOW) has been developed and the simulation results assured the schema efficiency.

J. Loureiro , O. Belo [9] have introduced an evolutionary approach of selection and allocation materialized view (sub cubes) over a distributed nodes. The solution utilizes a linear cost model, such that, it takes into account the communication costs and nodes processing power (e.g. aggregation, integration … etc.). The results have shown the superiority of genetic algorithms (GA) against the greedy algorithm in terms of cube selection and allocation as well as the scalability. Moreover, M. Lawrence et al [10] have proposed new query processing method over distributed nodes, the schema essentially rely on the Fragment Aggregation and Recombination (FAR) approach, in which the queries are divided into sub queries and sent to different nodes in order to be answered. The Site Brokers are the responsible servers for cache indexing, organization and coordination of resources within each site, the indexes, in turn, contain the data cube lattice structure as well as the R-Tree for each view indexing cached fragments. Implementation and evaluation have shown 50% - 60% response time reduction for realistic network parameters. However, the schema is lacked of mathematical models that might help designers in terms of which appropriate number of nodes should they utilize, and what is the best hierarchy in order to attain lowest response time. L. Wang has presented the most recent approach [11] in which followed mostly the same four layered structure that has been presented in [10]. Since the approach is specified to serve real estate investment systems, which narrows the assumptions; many simplifications have been assumed, such as the application service procedures are the same on each server and most critically the task assigning to be static which is mostly unpractical.

Among the two OLAP standards; F. Dehne and et al [12] have shown that MOLAP is suffering from scalability problems such that representing extremely sparse spaces is not easily adapted. On the other hand, ROLAP facing speed indexing problem for OLAP queries (performance problem). Accordingly, new solution has been proposed (RCUBE) based on maintaining the ROLAP scalability advantage with improving query indexing speed. The new approach uses packed R-trees, round robin distributed disk striping and Hilbert curve based (previous approaches used XYZ data format – also called lowX or nearest-X – which its response time deteriorate rapidly when relevant point are dispersed over the data set) for data ordering in order to achieve the minimum communication volume among p processors that process a balanced load and without a need for a shared disk in addition to maintain the scalability in terms of number of processors, data size and dimensions. The proposed solution considers that the queries are passed directly to each processor and the intermediate results remain distributed for further processing, previous RCUBE approaches considered global R-tree on the front-end. Furthermore, one of the significant improvements that have been presented; a threshold ($\alpha$) such that if the number of records in the result set is below the threshold, the partial result set are sent to one processor to be sorted, otherwise, will be sent to multiple processors, the $\alpha$ value can be set as a physical characteristics of the parallel machine.

The scheme has been implemented in C++, STL and MPI, and tested on a 17 nodes Beowulf cluster (a front-end and 16 compute nodes) with 1.8 GHz Intel Xeon processors, 1 GB RAM per node, 2X40 GB 7200 RPM IDE disk drives per node, the operating system was Linux Redhat 7.2 gcc 2.95.3 and all nodes are connected via 100 MB Ethernet switch. The results have shown that the index construction speed is near to the optimum, figure 1 shows some sample results:

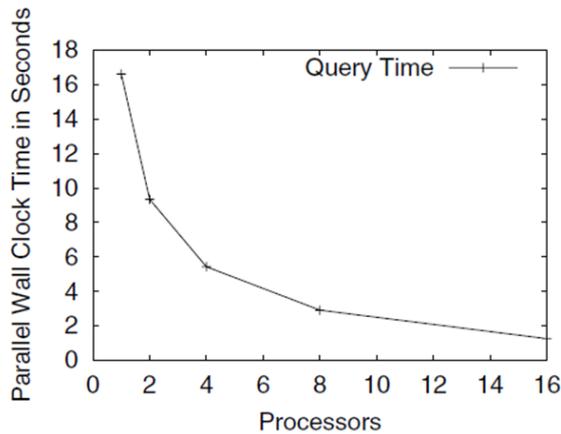 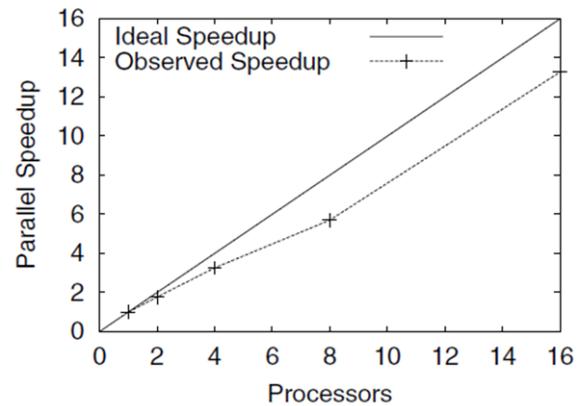

**Figure 1 (a).** Distributed query resolution [12].  **Figure 1 (b).** Corresponding speedup [12].

Another almost similar approach has been presented by J. You [13]. Namely, HDW solution, in which, GFS has been utilized as infrastructure for partitioning, load balance and failure tolerance purposes, Bigtable for distributed Cubetable management purposes and MapReduce for task parallelism purposes (The three components: GFS, Bigtable, MapReduce [3] pertain to Google distributed file system's infrastructure). Furthermore, MPI has been used for communication. The schema has been implemented in C++ and the conducted experiments in an 18-node PC cluster and 36 total cores, 18 GB RAM and 540 GB disk volumes, the experiments have shown that 60 million rows (1.37 GB text file) needed less than 5 minutes to be constructed. In addition, if the number of rows increases (e.g. 100 million) the construction time will not exceed 7 minutes. For the 36 nodes; the speedup is almost linear.

Regarding dataset distribution, there are two suggestions, first: data replication, second, centered dataset, the last suggestion has been indicated as undesired since it causes bottleneck in the central server. However, replication is considered as storage wasting choice also. An improvement has been suggested to be partial replication. In this regard there are many of what so called replica algorithms such as [14], in which, a new algorithm (OLAP4RS) has been presented based on pre-processing framework solution instead of post processing solutions (start processing the replica selection after creation request received).

In fact, in addition to the aforementioned examples; there are a significant number of researches that have addressed sequential and parallel cube construction methods (e.g. 15, 16, 17, and 18). However, all these approaches are following the static cube construction proposed by Gray et al [1]. The major disadvantage of such cubes is that the system can be only updated periodically. Therefore, the latest information might not be bundled with the relatively old ones in order to answer some queries. Another shortcoming is the massive space of memory that is needed due to materialized views storing in such structures as well as data duplicating.

## 3. Materialization Enhancements

The last issue – memory space - has been addressed by a considerable number of researches that suggest two solutions: compression and view selection.

### 3.1. Compression

The compression has been early suggested as a potential solution for enormous storage size. In the literature we can find a scalable and efficient parallel algorithm has been introduced for Closed Cube Computation – a lossless cube compression method – [19]. Utilizing MapReduce [3] framework for parallelism; the results have indicated significant reduction of storage size when the number of the closed cubes goes up. [21] Also has utilized MapReduce paradigm [3] for parallel closed cubes. Another example [20] by Z. Wang, Y, Xu have demonstrated the minimal condensed cubes challenges and discussed alternative computation algorithm (MinBST) that outperform previous ones. [22] has presented compression algorithm based on the balanced tree bitmap compression mechanism and has shown that the structure compression time can be logarithmic which outperforms the traditional bitmap compression techniques (linear time).

However, such a solution cannot be applied to the materialized views due to the fact that with more compression and storage size saving, there is relatively longer time needed for computation and answering the queries. Therefore, we can easily find in [20], [21], [22] the queries answer's time tends to rise which make the compression solution, as a result, inefficient for Real-Time purposes.

*3.2. View Selection*
Furthermore, some researches have focused on view selection issue, many algorithms have been suggested to choose the most suitable views to be loaded into the main memory due to the fact that we have enormous amount of data views and limited main memory size. Graph OLAP [23] is a suggested solution for views materialization issue by considering particular constrains in order to decide to choose partial or full materialization. A dynamic tuning of aggregation tables (materialized views) has been addressed as an online solution [24]. Aggregation tables tuning represent the stage of views selection in which an optimized selected views can be only loaded into the main memory to enhance the response performance. Recently the Genetic Algorithm (GA) has been introduced [25], [26] as one of potential solutions in this regard. CSSAA_VSP clonal based algorithm [27] has been proposed also to solve views selection issue. Remarkably, according to the empirical results, the algorithm has outperformed GA algorithm in terms of speed and solution quality, it has been proved as feasible and efficient algorithm. S. Sen et al [28] also have presented SelectPath algorithm that chooses the path which require generation of smallest number of cuboids among others. The approach utilizes Galois connection method. A numerical example has been given, yet the algorithm had not been investigated. Additionally, Galois connection has been used also in [29] to develop an algorithm utilizing dynamic data structure aiming at reducing time consuming and space occupation over time.

**4. Real-Time OLAP**
In order to bundle the latest updates with OLAP answers; researchers have focused on hardware processing capabilities. Three approaches have been addressed as follow:

*4.1. Multi-Core Processing Solution*
Authors F. Dehne and H. Zaboli [30] have presented a novel Parallel Real-time OLAP that is exploiting Multi-core processors systems. On contrast to the most introduced solutions in this regard (except rare instances such as [24] which utilize dynamic data cube); this approach does not rely upon views materialization or any static structure or pre-computed aggregations. Therefore, in order to compensate the performance degradation resulted from ignoring materialization solution; the authors have suggested taking advantage of the revolutionary hardware developments, specifically, multi-core processors that have the ability to handle several operations at the same time. In more details, the solution developed a parallel DC-tree for multi-core architectures as the sequential DC-tree [31] is the only fully dynamic data structure has been published so far. DC-tree is an extension of X-tree [32] and R-tree [33]. The new structure has been used with two main operations proposing two algorithms: 1) PARALLEL_OLAP_INSERT and 2) PARALLEL_OLAP_QUERY. The tests (using TPC_DS as benchmark) have demonstrated a significant speedup in terms of response time and throughput. A different approach to achieve Near Real-Time has been proposed by P. Mazur et al [34]. The solution suggests a universal database analysis repository model that contains all descriptive information about previous queries such that no need to repeat the whole calculation operation to find an answer, instead, small update can be applied to previous results to answer the query faster. Same concept has been portrayed also in [35] called the minimum incremental maintenance of materialized view, in which updating views occurring periodically and incrementally to avoid re-compute the entire view.

*4.2. GPU Solution*
I3DC (Interactive Three-Dimensional Cubes) is a substantial solution based entirely on GPU [36], the approach utilizes a new blending-as-aggregation (BAA) algorithm that maps OLAP aggregations to GPU mechanisms of rendering. Without the need of pre-computation and translation to integers by CPU; the new systems can build and visualize data cubes for tens of millions of records in milliseconds. The only concern of this approach is that show the results in chromatically graphical manner (in order to save up time of translation to integers and to be completely compatible with GPU structure and operations). Therefore, the output will be limited to the color range that the human eyes can recognize. However, the application that has been presented uses cubes of up to three dimensions, which is trivial number comparing to the real world datasets that can have much larger number of dimensions. Visualizing bigger numbers will definitely result in bigger number of colors which would cause difficulties for human eyes to distinguish between them.

*4.3. Multi-Core Processing + GPU Solution*
Exploiting the revolutionary hardware capabilities, M. Malik et al [4] have proposed three significant performance improvements: algorithm of string dictionary system used to translate literal data into integers for GPU processing purpose (translation performance evaluation based on TPC-DS fact tables and has shown 7% performance degrading due the translation time). Second, parallel version of OLAP processing using the CPU that has been implemented in OpenMP as well as a performance model needed by the scheduling algorithm in order to decide to which resource (CPU or GPU) should different queries be assigned. Third, the scheduling algorithm that supports text-to-integer translation and multiple CPU and GPU partitions. Each of the two resources has two main tasks; CPU processes OLAP queries that are not costing as well as performing text-to-integer translation. GPU in turns builds the cubes from traditional tables that are stored in GPU memory in addition to performing queries that are costly for CPU.

GPU has six partitions that are used to process queries from fact tables, while CPU has two partitions, figure 2 - [4]. The same idea (exploiting GPU to serve OLAP purposes) has been addressed also by C. Weyerhaeuser et al [38] with Business Intelligence Accelerator (BIA: highly distributed analytical engine supports OLAP). The results and comparison have shown substantial gain over CPU-based implementation although that the final recommendation was not to use GPU for BIA for particular reasons (e.g. GPU has smaller memory, data transfer from main memory to GPU memory has a cost … etc.)

## 5. Proposed Solution
Whatever improvements have been applied to materialization approach; none of them could overcome the Real-Time problem. Hence, our suggested approach is based on the concept of enhancing the hardware processing utilization. In particular, take advantage of both CPU and GPU. GPU significantly outperform CPU. However, using only GPU has shortcoming that the resultant answers will not be in numerical manner, rather, it might be multidimensional colourful graphs, shapes … etc. Hence, combining both CPU and GPU is a must in order to assign translation tasks to the CPU and processing tasks to the GPU. The proposed solution is essentially based on [4]. The algorithm will be based on TCP-DS benchmark by modifying the code of DsGen software which is open-source code. The software allows us to generate data with different scales from 100GB to 300TB. Afterward we can apply queries and compare the results in terms of response time. Our major contribution will be using this task partitions to achieve real-time OLAP, the following diagram shows the solution approach. Figure 2 shows the block diagrams of the suggested GPU and CPU partitions [4].

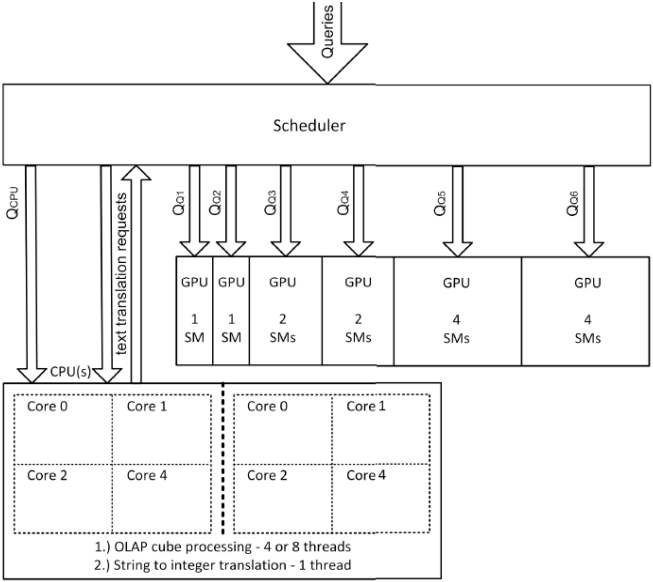

**Figure 2**. The block diagram of the GPU and CPU [4].

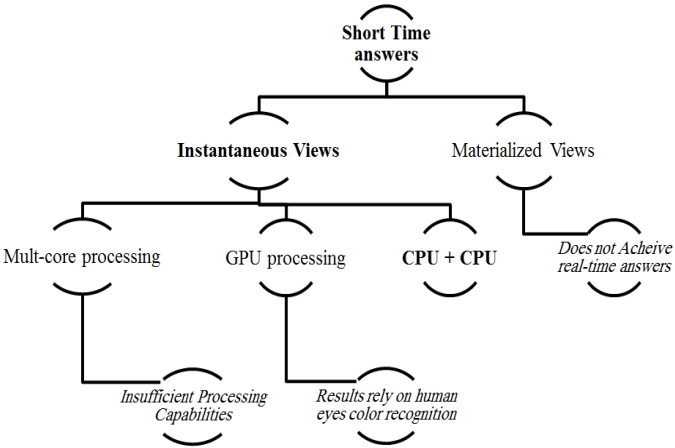

**Figure 3.** Graph of presented solutions for short time OLAP answers.

## 6. Expected Results
By applying the aforementioned approach we expect to achieve real-time answers by ignoring materialization approach and utilizing dynamic cubes rather than static ones. Furthermore, the increasing time needed to handle queries and processing them from the scratch will be compensated by exploiting the revolutionary processing capabilities that GPU offer. However, GPU by itself cannot present results in numerical manner, thus, including the multi-core processor is essential in order to accomplish two tasks, scheduling and translating data from literal to integer and vice versa. The approach tremendously affects the e-business [37] sector in addition to healthcare and artificial intelligence fields

## 7. Conclusion

In this framework paper essential approaches suggested by researches in order to optimize OLAP performance have been introduced on two levels: materialized answers and instantaneous answers. We have shown why materialized views cannot achieve real-time answer as well as why instantaneous answers are problematic issue. Hence, the powerful processing capabilities that are needed for instantaneous answers have been handled by GPU + CPU hybrid systems approach. Furthermore, an approach which utilizes a distributed algorithm presented in [12] and a partitioning algorithm presented in [4] in order to achieve real-time answers have been suggested. Future trends might implement the solution on Beowulf cluster and compare the enhancement in addition to developing the partitioning algorithm.

## Acknowledgments
This research work has been funded by MOHE Malaysia under research grant fund (RFGS 13-028-0269).